\documentclass[sigconf,screen]{acmart}
\AtBeginDocument{}

\usepackage{amsmath}
\usepackage[linesnumbered,ruled]{algorithm2e}
\usepackage{graphicx}
\usepackage{textcomp}
\usepackage{threeparttable}
\usepackage{tabularx}
\usepackage{booktabs}
\usepackage{multirow}
\usepackage{colortbl}
\usepackage{hyperref}
\usepackage{pifont}
\usepackage{bigstrut,multirow,rotating}
\usepackage{pifont}
\usepackage{tikz}

\iftrue
\usepackage{titlesec}
\titlespacing\section{0pt}{3pt plus 1pt minus 1pt}{0pt plus 1pt minus 1pt}
\titlespacing\subsection{0pt}{3pt plus 1pt minus 1pt}{0pt plus 1pt minus 1pt}
\titlespacing\subsubsection{0pt}{3pt plus 1pt minus 1pt}{2pt plus 1pt minus 1pt}
\fi

\copyrightyear{2024} 
\acmYear{2024} 
\setcopyright{acmlicensed}\acmConference[DAC '24]{61st ACM/IEEE Design Automation Conference}{June 23--27, 2024}{San Francisco, CA, USA}
\acmBooktitle{61st ACM/IEEE Design Automation Conference (DAC '24), June 23--27, 2024, San Francisco, CA, USA}
\acmDOI{10.1145/3649329.3655948}
\acmISBN{979-8-4007-0601-1/24/06}

\begin{document}



\title{Towards Efficient SRAM-PIM Architecture Design by Exploiting Unstructured Bit-Level Sparsity}

\titlenote{This work is supported in part by National Natural Science Foundation of China (Grant No. 62072019) and National Key Laboratory of Spintronics.
Corresponding authors are \textit{Jianlei Yang} and \textit{Weisheng Zhao}, Email: \url{jianlei@buaa.edu.cn}, \url{weisheng.zhao@buaa.edu.cn}}

\author{Cenlin Duan$^1$, \quad Jianlei Yang$^1$, \quad Yiou Wang$^1$, \quad Yikun Wang$^1$, \quad Yingjie Qi$^1$, }
\author{Xiaolin He$^1$, \quad Bonan Yan$^2$, \quad Xueyan Wang$^1$, \quad Xiaotao Jia$^1$, \quad Weisheng Zhao$^1$}

\affiliation{
    \institution{
        $^1$Beihang University, Beijing, China \hspace{3em}
        $^2$Peking University, Beijing, China \\
    }
    \country{}
}

 
\begin{abstract} \label{sec: Abstract}
Bit-level sparsity in neural network models harbors immense untapped potential. 
Eliminating redundant calculations of randomly distributed zero-bits significantly boosts computational efficiency. 
Yet, traditional digital SRAM-PIM architecture, limited by rigid crossbar architecture, struggles to effectively exploit this unstructured sparsity.
To address this challenge, we propose Dyadic Block PIM (DB-PIM), a groundbreaking algorithm-architecture co-design framework.
First, we propose an algorithm coupled with a distinctive sparsity pattern, termed a dyadic block (DB), that preserves the random distribution of non-zero bits to maintain accuracy while restricting the number of these bits in each weight to improve regularity.
Architecturally, we develop a custom PIM macro that includes dyadic block multiplication units (DBMUs) and Canonical Signed Digit (CSD)-based adder trees, specifically tailored for Multiply-Accumulate (MAC) operations.
An input pre-processing unit (IPU) further refines performance and efficiency by capitalizing on block-wise input sparsity.  
Results show that our proposed co-design framework achieves a remarkable speedup of up to $7.69\times$ and energy savings of $83.43\%$.

\end{abstract}

\setcopyright{none}

\maketitle
\pagestyle{plain}

\section{Introduction} \label{sec: Introduction}
Deep Neural Networks (DNNs) have become pervasive in numerous applications, including image recognition \cite{krizhevsky2012imagenet}, speech recognition \cite{zhang2017very}, and object detection \cite{yurtsever2020survey}.
However, the substantial memory and computing requisites pose challenges to performance and efficiency, especially for resource-constrained devices. 
Processing-in-memory (PIM), as an innovative computational paradigm, offers a potential solution to these challenges.
Distinct from the \textit{Von Neumann} architecture, PIM executes multiply-accumulate (MAC) operations in memory, thereby eliminating the \textit{Memory Wall} caused by frequent data movement.
Previous studies have shown that various technologies, such as SRAM \cite{chih202116,duan2023ddc}, RRAM \cite{yang2019sparse,liu2021bit}, and MRAM \cite{wang2020tcim,chen2022accelerating,zhao2023nand}, are available candidates for PIM. 
Among these technologies, SRAM is widely used in academia and industry due to its faster write speed, lower write energy, and compatibility with the most advanced process technologies.

Current SRAM-PIM research focuses on sparsity support to further improve computational efficiency. 
The majority of these studies have been geared towards leveraging value-level sparsity.
Bit-level sparsity, however, can further eliminate the redundant computation associated with zero bits in the values, which garnered significant interest.
It is worth noting that although bit-level sparsity has been explored and applied in traditional digital accelerators, its efficient utilization in SRAM-PIM still presents numerous challenges.

\begin{figure}[t]
  \centering
  \includegraphics[width=\linewidth]{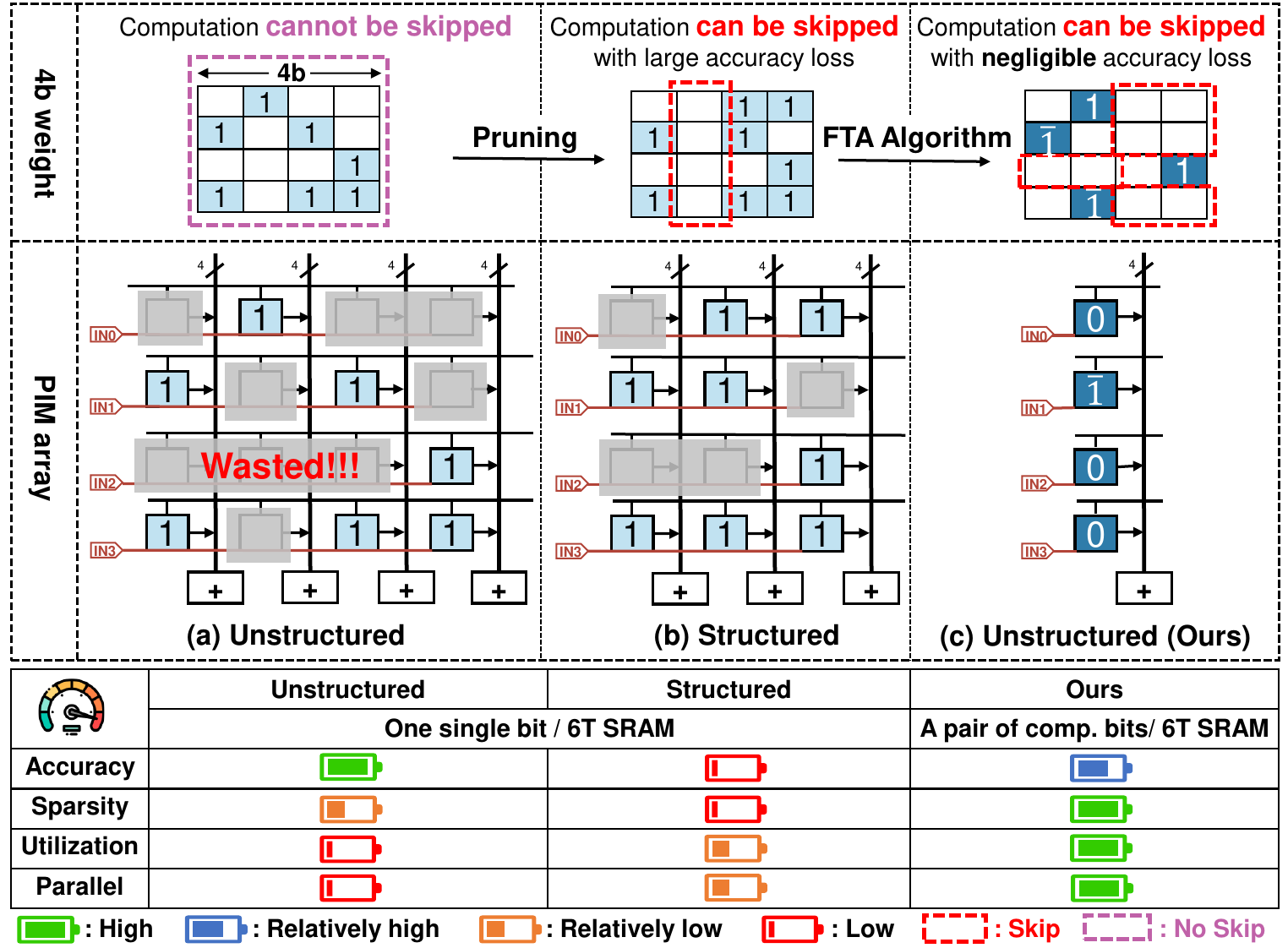}
  \vspace{-18pt}
  \caption{Exploitation of bit-level sparsity in SRAM-PIMs.}
\label{fig1:Comparison}
\vspace{-6pt}
\end{figure} 

\textbf{Computational dependency issues.} Unlike traditional digital accelerators, PIM-based accelerators are constrained by their rigid crossbar structure \cite{tu2022sdp}. 
This structure enforces strict data routing for input broadcast and vertical output accumulation, thereby impeding the efficient utilization of the randomly distributed zero bits.
As shown in Fig.~\ref{fig1:Comparison}(a), despite the high prevalence of zero bits, their random distribution prevents the system from efficiently bypassing them during computations, which dilutes the potential computational efficiency gains.
The challenge is particularly pronounced in the highly parallel architecture of digital SRAM-PIM, where the entire array is activated and computes concurrently.

\textbf{Low utilization issues.} To solve the above problem, one potential method is structured bit-level pruning, as shown in Fig.~\ref{fig1:Comparison}(b). 
However, this method eliminates only a minimal number of zero bits, concurrently resulting in significant accuracy degradation. 
As a result, the retention of zero-bit mapping within the PIM array results in a multitude of ineffectual computations. 
Despite various studies proposing refined mapping strategies to boost array utilization, pervasive bit-level sparsity still hinders optimal utilization.
To better quantify this issue, we define a particular \textit{actual utilization} by
\begin{equation}
\label{eq1}
\mathcal{U}_{act}=\frac{EffectiveCompSRAMCells}{TotalCompSRAMCells} \times 100\%.
\end{equation}
It represents the ratio of SRAM cells engaged in effective computation (computing non-zero bit) to the total SRAM cells currently involved in the computation. 
This metric helps assess the utilization efficiency of SRAM-PIM architectures in handling sparsity.
For instance, as illustrated in Fig.~\ref{fig1:Comparison}, the utilization rates are $7/16$ and $8/12$, respectively. 
In contrast, our approach achieves a $4/4$ ratio by using a 6T SRAM cell to store a pair of complementary (comp.) bits for parallelism, thus efficiently addressing the challenge of bit-level sparsity and enhancing computational efficiency.

We observe that the cross-coupled structure of 6T SRAM, along with in-memory customization features, provides a unique avenue for utilizing unstructured bit-level sparsity.
Based on this observation, we propose an efficient algorithm/architecture co-design framework to overcome the above two fundamental challenges.
The contributions of this work are as follows:

\begin{itemize}
\item We propose Dyadic Block PIM (DB-PIM), the first algorithm and architecture co-design framework tailored for digital SRAM-PIM that effectively harnesses the unstructured bit-level sparsity.
\item At the algorithm level, we present a Fixed Threshold Approximation (FTA) algorithm, alongside a unique bit-level sparsity pattern termed dyadic block (DB), both employing the Canonical Signed Digit (CSD) encoding method.
This approach maintains the random distribution of non-zero bits for accuracy while restricting the number of non-zero bits in each weight to improve regularity. 
It provides a foundation for solving computational dependency and low utilization issues in rigid crossbar structures.
\item At the architecture level, we design a customized PIM macro that incorporates dyadic block multiply units (DBMUs) and CSD-based adder trees, specifically designed for efficient MAC operations on randomly distributed non-zero bits.
Additionally, our architecture dynamically detects sparse blocks from input features, bypassing all-zero-bit blocks to enhance computational efficiency. 
\end{itemize}

The rest of this paper is organized as follows.
Section~\ref{sec: Motivation} provides background and motivation.
Section~\ref{sec: Methodology} demonstrates the details of the proposed methodology. 
Experimental results are illustrated in Section~\ref{sec: Experiments} and the conclusion is given in Section~\ref{sec: Conclusion}.

\section{Background and Motivation}
\label{sec: Motivation}
\subsection{Richness in Bit-level Sparsity}\label{sec:2.1}

\textbf{Limitation on Value-level Sparsity.} Recent studies demonstrate that a significant proportion of zero values exist within weights and input features of neural network (NN) models.
These zero values inherently contribute nothing to the final computational results. 
Thus, skipping the computations associated with these zero values can lead to considerable savings in computational resources, thereby boosting efficiency.  
However, the proportion of zero values in NN models is finite.
This indicates the full potential of computational efficiency gains, achievable by bypassing unnecessary computations, has not yet been completely tapped.
Consequently, research is increasingly shifting focus from the value-level sparsity to the finer granularity of bit-level sparsity. 
This transition not only opens new avenues for optimization but also aligns with the pressing need for more efficient computational paradigms.

\begin{figure}[t]
\centering
\includegraphics[width =\linewidth]{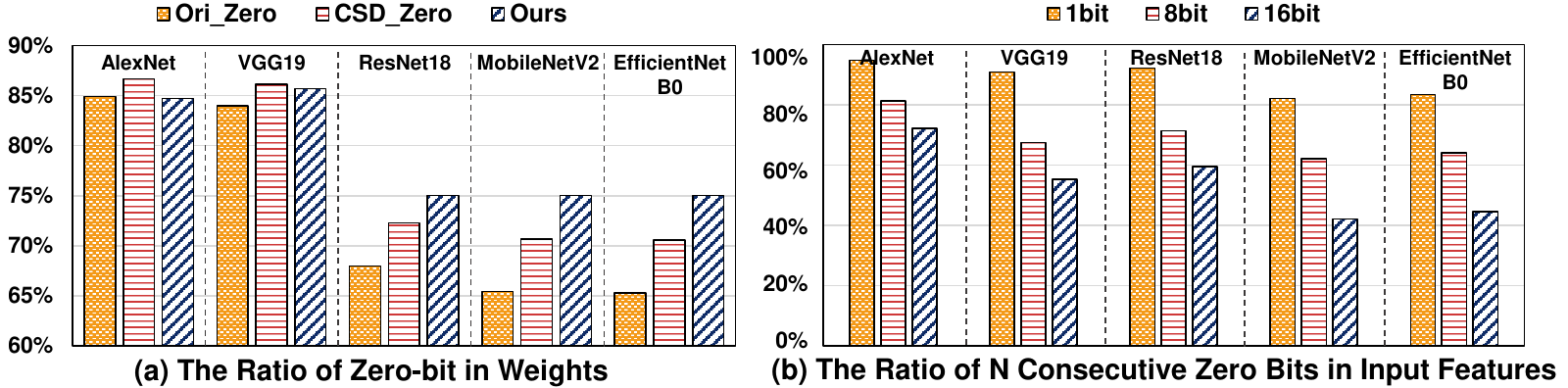}
\vspace{-18pt}
\caption{Bit-level sparsity existed in weights ($W$) and input features ($I$) among different models.}
\label{fig2:integration}
\end{figure}

\textbf{Opportunity on Bit-level Sparsity.} By delving into the bit-level sparsity within values, we uncover substantial untapped potential.
For example, the multiplication of \texttt{INT8} weight ($W$) and input feature ($I$) can be broken down into $64$ individual $1b \times 1b$ operations, as shown in Eq.~(\ref{eq2}).
Only those pairings where both $I_i$ and $W_j$ are non-zero effectively contribute to the final results.
\begin{equation}
\label{eq2}
I*W=\displaystyle\sum_{i=0}^7{\displaystyle\sum_{j=0}^7{I_{i} \times W_{j}}} \ .
\end{equation}
However, the proportion of non-zero bits in both weights and input features is relatively low, indicating that most of these computations are ineffectual. 
Fig.~\ref{fig2:integration}(a) illustrates the zero-bit proportion across various NN models.
Even when employing compact NN models such as MobileNetV2 and EfficientNetB0, bit-level sparsity still reaches a significant rate of approximately $65\%$.
Bypassing these redundant calculations can significantly enhance computational efficiency and open new avenues for model optimization.
To further exploit the potential of bit sparsity, CSD encoding—a distinctive digital representation is proposed, characterized by its prohibition of consecutive non-zero bits.
By employing CSD encoding, the overall sparsity is increased by an additional 5\% compared to the initial level of sparsity.
For compact models, our algorithm can improve by an additional $5\%$ based on CSD encoding.

Fig.~\ref{fig2:integration}(b) demonstrates bit-level sparsity in input features.
However, fully exploiting all non-zero bits is impractical due to substantial hardware overhead.
Our analysis indicates that when input features are grouped in sets of 8 or 16, the probability of identical bit positions being zero across the group is relatively high (up to $80\%$ in groups of $8$, and around $70\%$ in groups of $16$). 
Bypassing columns comprised entirely of zeros can yield efficiency gains in computation.  

\subsection{Sparsity Exploitation in SRAM-PIM}

\textbf{Value-level Sparsity.} Value-level sparsity has been extensively explored in SRAM-PIM architectures, as summarized in Tab.~\ref{tab:related works}.
For instance, `$W$ $+$ $V$' represents skipping multiplications where the \underline{w}eight \underline{v}alue is zero, while `$I$ $+$ $B$' represents skipping multiplications where the \underline{i}nput feature \underline{b}it is zero. 
Yue et al. \cite{yue202115} pioneered the implementation of block-wise zero-skipping in analog SRAM-PIM architectures, laying the groundwork for sparsity optimization. 
However, due to ADC limitation, only part of the cell array is activated simultaneously. 
To address this limitation, SDP \cite{tu2022sdp} proposes a novel digital SRAM-PIM with a double-broadcast hybrid-grained pruning method, activating all rows simultaneously for enhanced efficiency.
Building on these developments and to further mitigate accuracy loss,  Liu et al. \cite{liu202316} present a butterfly-network-based zero skipper for unstructured NN models.
However, efficiently exploiting such unstructured pruning techniques in PIM entails a significant increase in hardware overhead.
Despite the significant strides made by these pioneering studies, the prevalence of zero bits in values continues to hinder the full potential of efficiency and utilization in SRAM-PIM architectures.

\begin{table}[]
      \caption{Sparsity Exploitation Comparison among SRAM-PIMs.}
      \vspace{-8pt}
  \label{tab:related works}
  \resizebox{0.48\textwidth}{!}{
\begin{tabular}{|c|ccc|ccc|}
\hline
& \multicolumn{1}{c|}{\begin{tabular}[c]{@{}c@{}}Yue et al.\\ \cite{yue202115}\end{tabular}} & \multicolumn{1}{c|}{\begin{tabular}[c]{@{}c@{}}SDP \\ \cite{tu2022sdp}\end{tabular}} & \begin{tabular}[c]{@{}c@{}}Liu et al. \\ \cite{liu202316}\end{tabular} & \multicolumn{1}{c|}{\begin{tabular}[c]{@{}c@{}}Tu et al. \\ \cite{tu202316}\end{tabular}}          & \multicolumn{1}{c|}{\begin{tabular}[c]{@{}c@{}}TT@CIM \\ \cite{guo2022tt}\end{tabular}}             & Ours                                                                                  \\ \hline
Sparsity Type                                                         & \multicolumn{3}{c|}{Value ($V$)}                                                                                                                                                                                                             & \multicolumn{3}{c|}{Bit ($B$)}                                                                                                                                                                                                                                                            \\ \hline
Weight/Input ($W$/$I$)                                                       & \multicolumn{1}{c|}{$W$}                                                             & \multicolumn{1}{c|}{$W$}                                                       & $W$                                                              & \multicolumn{1}{c|}{$I$}                                                                      & \multicolumn{1}{c|}{$W$}                                                                      & $W$ + $I$                                                                                 \\ \hline
Digital/Analog ($D$/$A$)                                                    & \multicolumn{1}{c|}{$A$}                                                             & \multicolumn{1}{c|}{$D$}                                                       & $D$                                                              & \multicolumn{1}{c|}{$D$}                                                                      & \multicolumn{1}{c|}{$A$}                                                                      & $D$                                                                                     \\ \hline
Unstruct./Struct. ($U$/$S$) & \multicolumn{1}{c|}{$S$}                                                             & \multicolumn{1}{c|}{$S$}                                                       & $U$                                                             & \multicolumn{1}{c|}{$U$}                                                                      & \multicolumn{1}{c|}{$U$}                                                                      & $U$   \\ \hline
\begin{tabular}[c]{@{}c@{}}Ineffectual MAC \\ Removed\end{tabular}     & \multicolumn{1}{c|}{\begin{tabular}[c]{@{}c@{}}Zero \\ $W$ + $V$\end{tabular}}         & \multicolumn{1}{c|}{\begin{tabular}[c]{@{}c@{}}Zero \\ $W$ + $V$\end{tabular}}   & \begin{tabular}[c]{@{}c@{}}Zero \\ $W$ + $V$\end{tabular}          & \multicolumn{1}{c|}{\begin{tabular}[c]{@{}c@{}}Zero \\ $I$ + $B$\end{tabular}} & \multicolumn{1}{c|}{\begin{tabular}[c]{@{}c@{}}Zero \\ $W$ + $B$\end{tabular}}& \begin{tabular}[c]{@{}c@{}}Zero $W$ + $B$ and \\ Zero $I$ + $B$\end{tabular} \\ \hline
\end{tabular}
}
\vspace{-6pt}
\end{table}

\textbf{Bit-level Sparsity. }
In Sec.~\ref{sec:2.1}, we explore the substantial potential of bit-level sparsity across various NN models.
Recent research has increasingly focused on this domain. 
Tu et al. \cite{tu202316} developed the Bandwidth-Balanced CIM (BB-CIM) architecture to address computational imbalances caused by input bit-level sparsity, aiming to equalize input bit-width for improving efficiency.
Similarly, TT$@$CIM \cite{guo2022tt} seeks to enhance efficiency by analyzing and leveraging the higher sparsity ratio of zero bits in both two’s complement and one’s complement representations. 
Although numerous studies have attempted to explore the utilization of unstructured zero bits in digital SRAM-PIM, current methodologies still fall short of fully harnessing the potential of these randomly distributed zero bits.
As demonstrated in Fig.~\ref{fig1:Comparison}(a), the majority of zero bits still need to be stored and processed in SRAM-PIM, resulting in a substantial number of ineffective calculations and low utilization.
Fig.~\ref{fig1:Comparison}(b) illustrates that the structured bit-level pruning method can overcome the above limitations. 
Yet, for digital SRAM-PIM with high parallelism, a prerequisite is that a significant volume of bits must simultaneously be zero at identical locations.
Such a pruning method will introduce significant accuracy loss and does not eliminate all zero bits, impeding the achievement of optimal array utilization.

\textbf{Opportunity:} Presently, digital SRAM-PIM is predominantly optimized for limited unstructured or structured bit-level sparsity, yielding only slight improvements. 
To bridge this gap, we propose a DB-PIM framework.
It aims to comprehensively exploit unstructured bit-level sparsity, signifying a paradigm shift in PIM design.
\textbf{By CSD encoding, FTA algorithm, and DB-PIM architecture, we selectively store and compute effective non-zero bits.
This approach significantly reduces invalid computations, enhancing efficiency, and simultaneously achieves considerably higher utilization.}

\section{Methodology}\label{sec: Methodology}
\subsection{Overall Framework}
\begin{figure}[t]
\centering
\includegraphics[width=\linewidth]{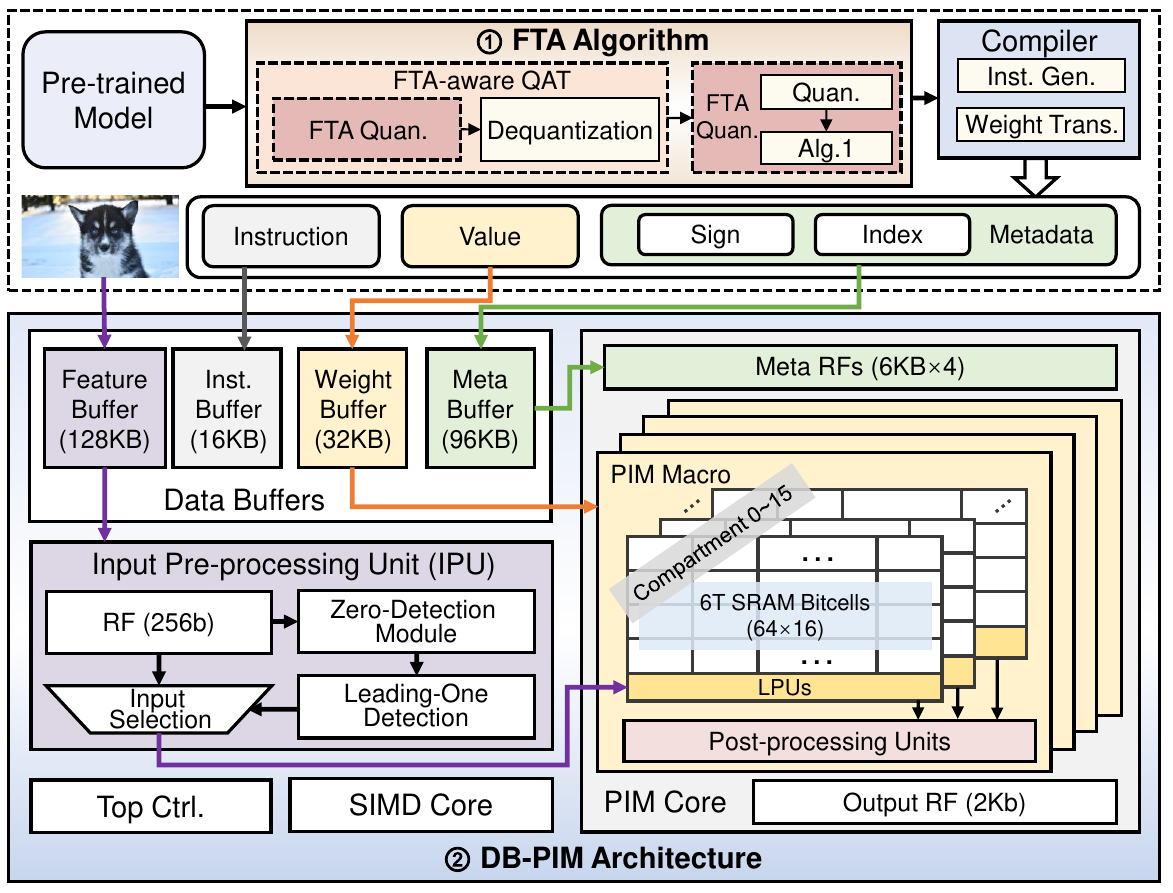}
\vspace{-16pt}
\caption{Overview of the proposed DB-PIM, an algorithm and architecture co-design framework.}
\label{fig3:Overview}
\end{figure}

Fig.~\ref{fig3:Overview} illustrates the overview of the DB-PIM, an algorithm and architecture co-design framework that effectively resolves the challenges discussed earlier. 
We first leverage the \textcircled{1} FTA algorithm and a bit-level sparsity pattern to obtain a weight matrix with a fixed number of non-zero bits. 
This approach maintains model accuracy by preserving the randomness of the non-zero bits and enhances data regularity by limiting the number of non-zero bits in each weight.
It forms the foundation for our framework to handle unstructured non-zero bits efficiently.
Subsequently, we develop a dedicated \textcircled{2} DB-PIM architecture, designed to accelerate these patterns. 
This cohesive integration of the FTA algorithm with the DB-PIM architecture effectively addresses computational dependencies typically found in rigid crossbar structures, enabling efficient computation of randomly distributed non-zero bits.
Consequently, this enhances hardware efficiency and optimizes array utilization.
Specifically, in the training procedure, we first apply a modified Quantization-Aware Training (QAT), known as FTA-aware QAT, to the pre-trained model.
This step is crucial for obtaining quantization parameters that reflect the impact of our algorithm on model accuracy.
Following this, we perform FTA quantization to obtain approximation models based on our FTA algorithm.
We then transform these models into values and metadata (including signs and indices) and generate corresponding instructions in the compilation phase. 
This compilation is conducted offline and stores the above information in off-chip memory.
Finally, DB-PIM performs bitwise \texttt{AND} operations with these values and recovers the final results based on the metadata in the customized PIM macros.

\subsection{FTA Algorithm}

\begin{figure*}[t]
\centering
\includegraphics[width =\linewidth]{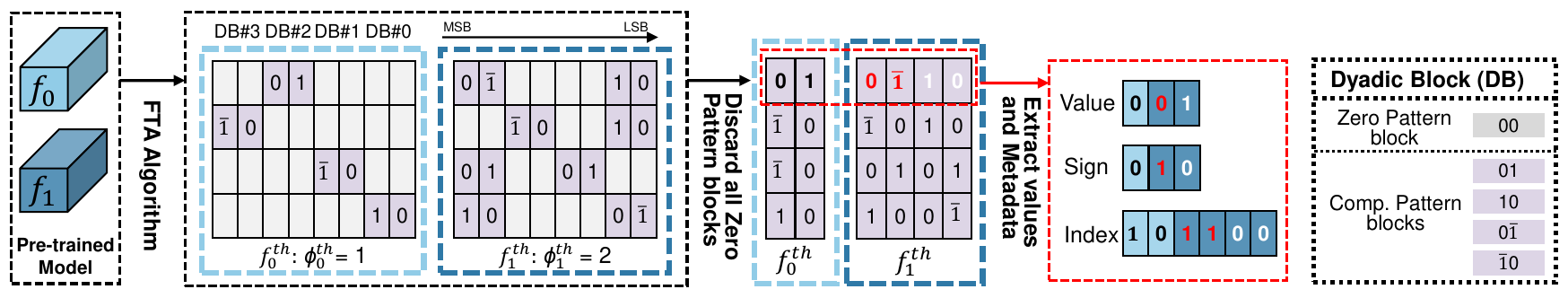}
\vspace{-16pt}
\caption{Extraction and representation of bit-level sparsity patterns.}
\label{fig4:Algorithm}
\vspace{-6pt}
\end{figure*}

\textbf{CSD Representation for Weight. } The CSD encoding, utilized in our algorithm, is a binary number representation system that employs three possible digit values: $1$, $0$, or $-1$ (with $-1$ often represented by $\overline{1}$). 
A hallmark characteristic of CSD is the constraint that adjacent bits cannot be both non-zero bits ($1$, $\overline{1}$).
This restriction ensures each number has a unique representation with a minimized count of non-zero bits. 
On average, the CSD representation contains 33\% fewer non-zero bits than their binary equivalents, enhancing its bit-level sparsity and potentially reducing computational overhead \cite{pinjare2013implementation}. 
For example, the binary number $0111\_1101_2$ would be encoded in CSD as $1000\_0\overline{1}01_{CSD}$. 
Incorporating CSD encoding into our framework is driven by two primary factors.
First, CSD encoding significantly improves bit-level sparsity, potentially improving computational efficiency.
Secondly, its inherent property of preventing the occurrence of consecutive non-zero bits is crucial for our DB-PIM architecture.
This attribute aligns seamlessly with the hardware constraints and optimizes the architecture's operational efficiency, making it an ideal choice for our DB-PIM framework.

\textbf{Bit-level Sparsity Pattern.} Based on CSD encoding, we propose a novel bit-level sparsity pattern termed the dyadic block (DB). 
This pattern partitions an 8-bit binary number into four DBs, each consisting of a pair of bits. 
As a fundamental unit of our encoding scheme, the DB is allocated a distinctive index to denote its position, which is crucial for accurately locating and processing non-zero bits.
In our CSD encoding, the DB can be divided into two categories: Zero Pattern block ($00$) and Complementary Pattern (Comp. Pattern) blocks, which include $01, 10, 0\overline{1}$, and $\overline{1}0$.
To effectively quantify sparsity within our framework, we introduce a symbol $\phi$ to represent the number of non-zero bits, which ranges from 0 to 4 and corresponds to sparsity levels from 100\% down to 50\%. 
As depicted in Fig.~\ref{fig4:Algorithm}, the 8-bit binary number $f_1^{th}(0) = 0\overline{1}00\_0010_{CSD}$ decomposes into four DBs: `$0\overline{1}|00|00|10$', assigning each pair an index, `$DB\#3|DB\#2|DB\#1|DB\#0$'. 
$\phi_1(0) = 2$ reflects the two non-zero bits in this value and 75\% sparsity. 

\begin{algorithm}[t]
\small
\KwIn{Quantized filters ${\mathcal{F}} \doteq \left[ f_0, \dots, f_{n-1} \right]$, where $f_{i} \in  D^{\mathcal{N}}$, $n$ is the number of filters, $\mathcal{N}$ is the number of elements in one filter, $D$ is determined by the quantization precision, e.g. \texttt{INT8}, Query Table $T(\phi^{th}) =\left \{ t \in D\mid \phi(toCSD(t)) = \phi_i^{th}\right \}$}
\KwOut{Approximation filters ${\mathcal{F}}^{th} \doteq \left[ f^{th}_0,\dots,f^{th}_{n-1} \right]$, filter thresholds $\Phi^{th} \doteq \left[ \phi^{th}_0,\dots,\phi^{th}_{n-1} \right]$.}
    \For{i \textbf{in} $[ 0, 1, 2, \dots,  n-1 ]$}{
        \For{j \textbf{in} $[0, 1, 2, \dots, {\mathcal{N}} - 1]$ }{
            $f^{csd}_i(j) \gets \text{toCSD}(f_{i}(j))$ \hfill{\textcolor{brown}{// CSD Conversion}}\\
            $\phi_i(j) \gets \text{CountNonZeros}(f^{csd}_{i}(j))$ \hfill{\textcolor{brown}{//Count Non-Zero Bit}} 
        }
        $m_i \gets Mode\left \{ 0\le j < \mathcal{N}\mid \phi_i(j)\right \}$ \hfill{\textcolor{brown}{//Compute Mode}}\\
        \uIf{$\forall j, \phi_i(j) ==0 $}{
            $\phi_i^{th} \gets 0$  \hfill{\textcolor{brown}{// All Zero Filter}} \\ 
           }
        \uElseIf{$m_i == 0$}{
           $\phi_i^{th} \gets 1$ \\
           }
        \uElseIf{$ 1\le m_i \le 2$}{
            $\phi_i^{th} \gets m_i$ \\
        }
        \uElseIf{$ m_i >  2$}{
            $\phi_i^{th} \gets 2$  \hfill{\textcolor{brown}{// Limit Max Threshold to 2}}
        }
        \For{j \textbf{in} $[0, 1, 2, \dots, {\mathcal{N}} - 1]$}{
           $f^{th}_i(j) \gets \underset {t \in T(\phi^{th}) } {\arg\min}\left | t - f_i(j) \right | $  \hfill{\textcolor{brown}{//  Closest Num to $f_i(j)$ in $T$}}\\
    }
    }
\caption{Fixed Threshold ($\Phi_{th}$) Approximation (FTA)}
\label{alg1}
\end{algorithm}

The rationale for adopting this pattern is our discovery of high compatibility between the Comp. Pattern block and the cross-coupled structure in 6T SRAM cell. 
Specifically, a Comp. Pattern block can be stored within a single 6T SRAM cell and computed simultaneously. 
For each 8-bit value, we could eliminate all Zero Pattern blocks and only store and compute the Comp. Pattern blocks.
This approach not only preserves the unstructured characteristic of bit-level sparsity but also capitalizes on the cross-coupled structure to ensure the utilization efficiency of the SRAM cell array. 
In this system, all SRAM cells involved in the computation are effectively utilized, addressing \textbf{the low utilization issues}, as shown in Fig.~\ref{fig1:Comparison}.
However, merely exploiting this pattern falls short of resolving \textbf{the computational dependency issues}.
The reason lies in the variability of $\phi$ values across the weights in each filter. 
Indiscriminate removal of all Zero Pattern blocks might result in computational irregularity.
Such irregularity conflicts with the structured computation requirement of a rigid crossbar architecture.
To handle the computational dependency issues effectively, we propose the FTA algorithm. 

\textbf{Detailed Procedure of FTA Algorithm.}
The core idea of our algorithm lies in setting a uniform threshold, denoted as $\phi^{th}$ for each filter.
This threshold is pivotal in ensuring that weights in each filter maintain a uniform count of non-zero bits. 
Despite the random distribution of these non-zero bits, this consistency guarantees the elimination of the same number of Zero Pattern blocks.
As a result, it upholds the regular structure of each compressed filter, as shown in Fig.~\ref{fig4:Algorithm}. 
Specifically, the processing of this algorithm is shown in Alg.~\ref{alg1}.
First, the entire layer is grouped based on the number of filters, with the weights in these groups subsequently converted into CSD representation. 
Next, we determine a threshold $\phi^{th}$ for each filter by analyzing the distribution of the number of non-zero bits across all weights within the filter.
Weight distribution analysis in various NN models reveals that a $\phi^{th}$ value of 2 is the most prevalent across the filters. 
Therefore, to enhance sparsity while avoiding a significant accuracy drop, we confine the $\phi^{th}$ from $0$ to $2$.
The FTA algorithm sets the value of $f^{th}_i(j)$ as the closest value to $f_i(j)$ in set $T(\phi^{th})$. 
In conclusion, combining the FTA algorithm and sparsity pattern lays the foundation for addressing the prevalent computational dependencies and low utilization issues in SRAM-PIM. 

Our proposed FTA algorithm demonstrates significant potential in alleviating the bit-level irregularity by setting the uniformed $\phi^{th}$.
However, this potential remains underutilized in current SRAM-PIM architectures due to several limitations. 
First, the current PIM macro cannot execute parallel computations on the complementary states, denoted as $Q/\overline{Q}$,  stored in the cross-coupled structure of 6T SRAM.
Second, the existing adder trees cannot directly accumulate outputs with randomly distributed non-zero bits.
Third, the current PIM macro does not have the functionality to identify and bypass zero bits in input features during runtime.
As such, the DB-PIM architecture is specifically designed to capitalize on the novel opportunities offered by the FTA algorithm to enhance computation efficiency.

\subsection{Architecture Design of DB-PIM}
\textbf{Top Level Architecture.} Fig.~\ref{fig3:Overview} \textcircled{2} illustrates the overall architecture of DB-PIM, which is composed of a top controller, an input pre-processing unit (IPU), data buffers, a PIM core, and a SIMD core. 
The PIM core consists of metadata register files (RFs) for storing sign and index information fetched from meta buffer, four customized PIM macros, and an output RF.
The PIM macro is an extension of the ADC-less SRAM PIM macro proposed in \cite{yan20221}.
The top controller first processes instructions fetched from the instruction buffer (Inst. Buffer) and dispatches corresponding control signals to the whole system. 
The input features, stored in the feature buffer, are accessed by IPU for converting into bit-serial form.
IPU identifies and eliminates all-zero sequences, subsequently broadcasting all non-zero sequences along with their index information to the PIM core.
The PIM core, receiving weights from the weight buffer and input features from the IPU, executes bitwise \texttt{AND} operations.
Subsequently, the results of these operations are accumulated in a customized CSD-based adder tree, guided by metadata derived from the meta RFs.
These MAC results are then shifted and accumulated based on their respective bit position obtained from IPU for producing the partial sum (Psum). 
Then, we accumulate Psum, and the final results are written back in the feature buffer. 
The SIMD core is responsible for performing other element-wise operations.

\begin{figure}[t]
\centering
\includegraphics[width =\linewidth]{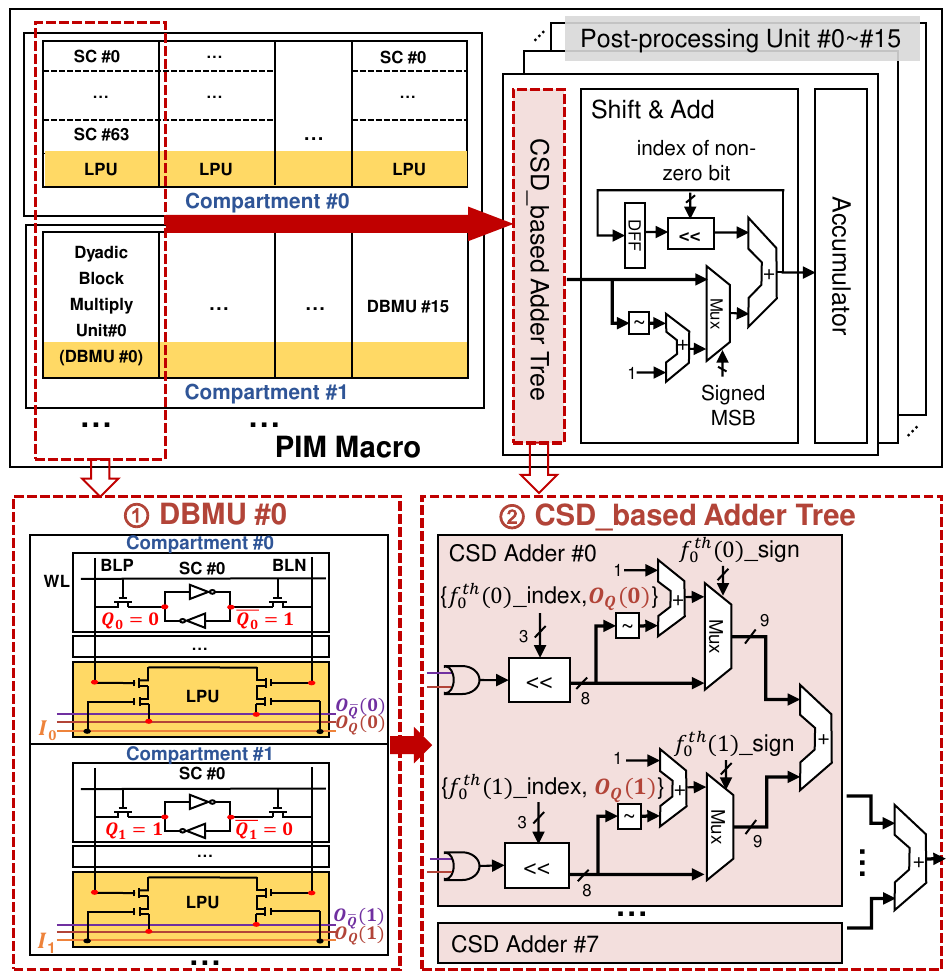}
\vspace{-16pt}
\caption{Circuit design of customized SRAM-PIM macro.}
\label{fig5:post-processing}
\vspace{-6pt}
\end{figure}

\textbf{Customized SRAM-PIM Macro.} Fig.~\ref{fig5:post-processing} showcases the circuit design of our customized SRAM-PIM macro, which consists of $16$ compartments, $16$ post-processing units, and other peripheral circuits. 
Each compartment comprises $16$ DBMUs, including sixty-four 6T SRAM cells (SC \#0 $\sim$ SC \#63) and one local processing unit (LPU).
The allocation of weights per row in each compartment is determined by both compartment width and the $\phi^{th}$ of each filter, equating to $8$ for $\phi^{th} = 2$ and $16$ for $\phi^{th} = 1$. 
In this setup, each LPU within a DBMU acts as a fundamental dyadic block multiplier, utilizing four transistors for two independent multiplications, $IN\times Q$ and $IN\times \overline{Q}$. 
Specifically, a Comp. Pattern block, stored in a cross-coupled structure ($Q$ and $\overline{Q}$) within the 6T SRAM cell of DBMU, executes two individual bitwise \texttt{AND} operations with identical input. 

Consider $f_0^{th}(0) = 0001\_0000_{CSD}$ and $f_0^{th}(1) = \overline{1}000\_0000_{CSD}$, as shown in Fig.~\ref{fig4:Algorithm} and Fig.~\ref{fig5:post-processing}. 
We eliminate all Zero Pattern blocks and keep Comp. Pattern blocks with their corresponding indices and signs. 
Then, we store $01$ (DB $\#2$ of $f_0^{th}(0)$) into $Q_0 /\overline{Q_0}$ with $sign = 0$ and $index = 10$, and $10$ (DB $\#3$ of $f_0^{th}(1)$) into $Q_1$ /$\overline{Q_1}$ with $sign = 1$ and $index = 11$ in Fig.~\ref{fig5:post-processing} \textcircled{1} DBMU $\#0$.
Subsequently, $I_0$ and $I_1$ are sent to Compartment $\#0$ and $\#1$ for bitwise \texttt{AND} operations.
$O_Q(0) = Q_0 \& I_0$, $O_{\overline{Q}}(0) = \overline{Q}_0 \& I_0$, $O_Q(1) = Q_1 \& I_1$, $O_{\overline{Q}}(1) = \overline{Q}_1 \& I_1$.
However, it is noteworthy that directly adding $\{O_Q(0),O_{\overline{Q}}(0)\}$ and $\{O_Q(1),O_{\overline{Q}}(1)\}$ would yield an incorrect results.
For instance, if $I_0$ and $I_1$ are both $1$, the sum would incorrectly be $11$.  
However, the correct result of $f_0^{th}(0) + f_0^{th}(1) = 0001\_0000_{CSD} + \overline{1}000\_0000_{CSD} =1\_1001\_0000_2$.
Thus, we have engineered a specialized \textcircled{2} CSD\_based adder tree, adept at handling randomly distributed non-zero bits. 
This innovation, integrated into our customized PIM macro and combined with our algorithm, allows us to exploit unstructured bit-level sparsity in digital SRAM.

\begin{figure}[t]
\centering
\includegraphics[width =0.9\linewidth]{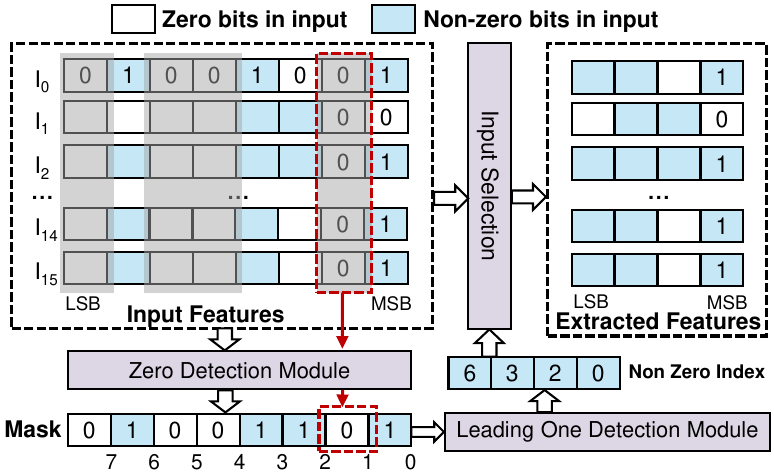}
\vspace{-10pt}
\caption{Bit-level sparsity utilization for input features.}
\label{fig6:IPU}
\vspace{-6pt}
\end{figure} 

\textbf{Input Pre-processing Unit.} Recognizing that the number of non-zero bits varies across different input features, it is impractical to bypass all zero bits directly within a rigid crossbar structure.
However, the data analysis presented in Fig.~\ref{fig2:integration}(b) reveals that block-wise zero bits still account for a large proportion.
To harness the bit sparsity in the input feature, we propose IPU, designed to dynamically detect block-wise zero bits, as shown in Fig.~\ref{fig6:IPU}.
Initially, the IPU identifies columns consisting entirely of zero bits and generates a corresponding mask.
Then, we detect the first non-zero bit in the above mask to select input features for calculations, thereby enhancing computational efficiency.

\section{Evaluation Results}\label{sec: Experiments}

\begin{table}
      \caption{Top-1 Accuracy Comparison on CIFAR100 Dateset.}
      \vspace{-8pt}
  \label{tab:accuracy}
  \resizebox{0.48\textwidth}{!}{
  \begin{tabular}{lrrrr}
    \toprule
\multicolumn{1}{l}{Models}    & \multicolumn{1}{c}{$W$/$I$ Precision} 
                              & \multicolumn{1}{c}{Ori. Accu.} 
                              & \multicolumn{1}{c}{FTA Accu.} 
                              & \multicolumn{1}{c}{Accu. Drop} \\
    \midrule
    AlexNet         &      8b/8b      &       70.62\%               & 69.64\%     & 0.98\%$\downarrow$     \\
    VGG19           &      8b/8b      &       76.78\%               & 76.14\%     & 0.64\%$\downarrow$     \\
    ResNet18        &      8b/8b      &       79.65\%               & 79.09\%     & 0.56\%$\downarrow$     \\
    MobileNetV2     &      8b/8b      &       82.20\%               & 82.04\%     & 0.16\%$\downarrow$     \\
    EfficientNetB0 &      8b/8b      &       72.19\%               & 71.67\%     & 0.52\%$\downarrow$     \\
    \bottomrule
  \end{tabular}
  }
  \vspace{-18pt}
\end{table}

\begin{table*}[]
\caption{Detailed Comparisons with Related Works.}
\vspace{-8pt}
\label{tab:comparison}
\resizebox{\textwidth}{!}{
\begin{tabular}{|cc|c|c|c|c|c|ccccc|}
\hline
\multicolumn{2}{|c|}{}                       & \cite{yue202115}   & \cite{tu2022sdp}  & \cite{liu202316}  & \cite{tu202316}  & \cite{guo2022tt}  & \multicolumn{5}{c|}{This Work}  \\ \hline
\multicolumn{2}{|c|}{Technology (nm)}        & 65                 & 28                & 28                & 28               & 28                & \multicolumn{5}{c|}{28}   \\ \hline
\multicolumn{2}{|c|}{Die Area (mm$^2$)}      & 12                 & 6.07              & 3.93              & 14.36            & 8.97              & \multicolumn{5}{c|}{1.15}  \\ \hline
\multicolumn{2}{|c|}{Supply Voltage (V)}     & 0.62$\sim$1        & 1                 & 0.64$\sim$1.03    & 0.60$\sim$1     & 0.60$\sim$0.90      & \multicolumn{5}{c|}{0.72$\sim$0.90}     \\ \hline
\multicolumn{2}{|c|}{Frequency   (MHz)}      & 25$\sim$100        & 500               & 20$\sim$320       & 85$\sim$275      & 125$\sim$216      & \multicolumn{5}{c|}{500}      \\ \hline
\multicolumn{2}{|c|}{Power (mW)}             & 18.60$\sim$84.10     & 1050              & 8.27$\sim$250.65  & 29.83$\sim$153.62& 11.40$\sim$45.10    & \multicolumn{5}{c|}{1.45$\sim$11.65} \\ \hline
\multicolumn{2}{|c|}{SRAM Size (KB)}         & 294                & 384               & 96                & 192              & 114               & \multicolumn{5}{c|}{272}  \\ \hline
\multicolumn{2}{|c|}{PIM Size (KB)}          & 8                  & 128               & 144               & 128              & 128               & \multicolumn{5}{c|}{8}  \\ \hline
\multicolumn{2}{|c|}{Number of PIM Macro}    & 4                  & 512               & 96                & 128              & 16                & \multicolumn{5}{c|}{4}  \\ \hline
\multicolumn{2}{|c|}{Dataset}                & CIFAR10 /ImageNet           & ImageNet          & Enwik8            & VQA              & CIFAR10          & \multicolumn{5}{c|}{CIFAR100}   \\ \hline
\multicolumn{2}{|c|}{\multirow{2}{*}{\textbf{Actual Utilization} ($\mathcal{U}_{act}$)}}                                                    & ResNet18           & ResNet-50                   & \multicolumn{1}{c|}{\begin{tabular}[c]{@{}c@{}}Adaptive-Span\\Transformer\end{tabular} }       & \multicolumn{1}{c|}{\begin{tabular}[c]{@{}c@{}}ViLBERT-\\base\end{tabular}}                                                                 & \multicolumn{1}{c|}{\begin{tabular}[c]{@{}c@{}}ResNet20\end{tabular} } & \multicolumn{1}{c|}{AlexNet}      & \multicolumn{1}{c|}{VGG19}      & \multicolumn{1}{c|}{ResNet18}        & \multicolumn{1}{c|}{MobileNetV2}             &       EfficientNetB0                \\ \cline{3-12} 
\multicolumn{2}{|c|}{}   &  \textbf{32.04\%} 
                                                     & \textbf{48.64\%} 
                                                     & \textbf{/}        
                                                     & \textbf{/}       
                                                     & \textbf{<50\%}   
                                                     & \multicolumn{1}{c|}{\textbf{91.95\%}} 
                                                     & \multicolumn{1}{c|}{\textbf{97.69\%}} 
                                                     & \multicolumn{1}{c|}{\textbf{98.42\%}}        
                                                     & \multicolumn{1}{c|}{\textbf{97.82\%}}   
                                                     & \textbf{94.41\%}      \\ \hline
\multicolumn{2}{|c|}{\textbf{Peak Throughput (TOPS) (8b/8b)}}   & \textbf{0.10 }        
                                                               & \textbf{26.21 }  
                                                               & \textbf{3.33  }     
                                                               & \textbf{3.55  }  
                                                               & \textbf{0.40  }  
                                                               & \multicolumn{5}{c|}{\textbf{0.31 }}      \\ \hline
\multicolumn{2}{|c|}{\textbf{\begin{tabular}[c]{@{}c@{}}Peak Throughput/Macro (GOPS) (8b/8b)\end{tabular}}}  & \textbf{24.69 } 
                                                                                                     & \textbf{51.19 }        
                                                                                                     & \textbf{34.68 }        
                                                                                                     & \textbf{27.73 }                                                               
                                                                                                     & \textbf{25.1 }       
                                                                                                     & \multicolumn{5}{c|}{\textbf{77.5}}   \\ \hline
\multicolumn{2}{|c|}{\textbf{\begin{tabular}[c]{@{}c@{}} Energy Efficiency (TOPS/W) (8b/8b)\end{tabular}}}              & \textbf{0.09$\sim$2.37}      
                                                                                                                      & \textbf{25 (dense)$\sim$107.60} 
                                                                                                                      & \textbf{1.96$\sim$25.22 }         
                                                                                                                      & \textbf{48.40$\sim$101 } 
                                                                                                                      & \textbf{5.99$\sim$13.75 } 
                                                                                                                      & \multicolumn{5}{c|}{\textbf{18.14$\sim$45.20 }}   \\ \hline
\multicolumn{2}{|c|}{\textbf{\begin{tabular}[c]{@{}c@{}} Peak Energy Efficiency per Unit Area \\ (TOPS/W/mm$^2$) (8b/8b)\end{tabular}}}              & \textbf{2.97}      
                                                                                                                      & \textbf{17.73} 
                                                                                                                      & \textbf{6.42 }         
                                                                                                                      & \textbf{7.03} 
                                                                                                                      & \textbf{1.53 } 
                                                                                                                      & \multicolumn{5}{c|}{\textbf{39.30}}   \\ \hline
\end{tabular}
}
\vspace{-8pt}
\end{table*}

\subsection{Experimental Setup}
\textbf{Hardware Implementation.} DB-PIM is evaluated on $28$ nm technology, with a $128$ KB feature buffer, a $16$ KB instruction buffer, a $32$ KB weight buffer, a $96$ KB meta buffer, four $6$ KB meta RFs and four $16$ Kb PIM macros. 
The power consumption, latency, and area of PIM macros are extracted from the post-layout of customized design extension from \cite{yan20221}.
The remaining digital modules are implemented with Verilog HDL and synthesized by Design Compiler for area evaluation, while PrimeTime PX obtains power consumption. 
Aiming to evaluate the performance of DB-PIM, we implement a simple compilation tool for dataflow mapping and a customized cycle-accurate C++ simulator for functionalities validation.

\textbf{Dense Digital PIM Baseline.} To evaluate the benefits and associated overhead of our proposed techniques, we establish a dense digital PIM baseline for comparison. 
This baseline, obtained by removing all the sparsity support from the DB-PIM architecture, consists of buffers, the SIMD core, PIM macros, an output RF, and IPU. 
Within this baseline, the PIM macro is similar to the state-of-the-art digital PIM \cite{yan20221}, while all other hardware configurations are the same as in the DB-PIM.

\subsection{Evaluation of FTA Algorithm}\label{sec: Evaluation of FTA Algorithm}
To demonstrate the generality of the FTA algorithm, we conduct evaluations across various NN models on the CIFAR100 dataset.
These models include standard NN models such as AlexNet \cite{krizhevsky2012imagenet}, VGG-19 \cite{simonyan2014very} and ResNet-18 \cite{he2016deep}, along with compact NN models such as MobileNetV2 \cite{sandler2018mobilenetv2} and EfficientNetB0 \cite{tan2019efficientnet}.
As detailed in previous research \cite{tu2022sdp}, compact NN models exhibit markedly less redundancy compared to standard models.
This inherent attribute suggests that traditional value-level sparsity methods are effective only on certain layers to maintain accuracy.
Thus, these methods, while effective in standard NN models, do not yield significant acceleration when applied to compact NN models.
Conversely, Tab.~\ref{tab:accuracy} details the accuracy loss associated with our algorithm.
Importantly, the accuracy loss remains below $1\%$ even when applying the FTA algorithm to all layers of compact NN models.
This underscores the effectiveness of our algorithm in various NN models with negligible accuracy loss.


\subsection{Evaluation for Hybrid Sparsity}

\begin{figure}[t]
\centering
\includegraphics[width =\linewidth]{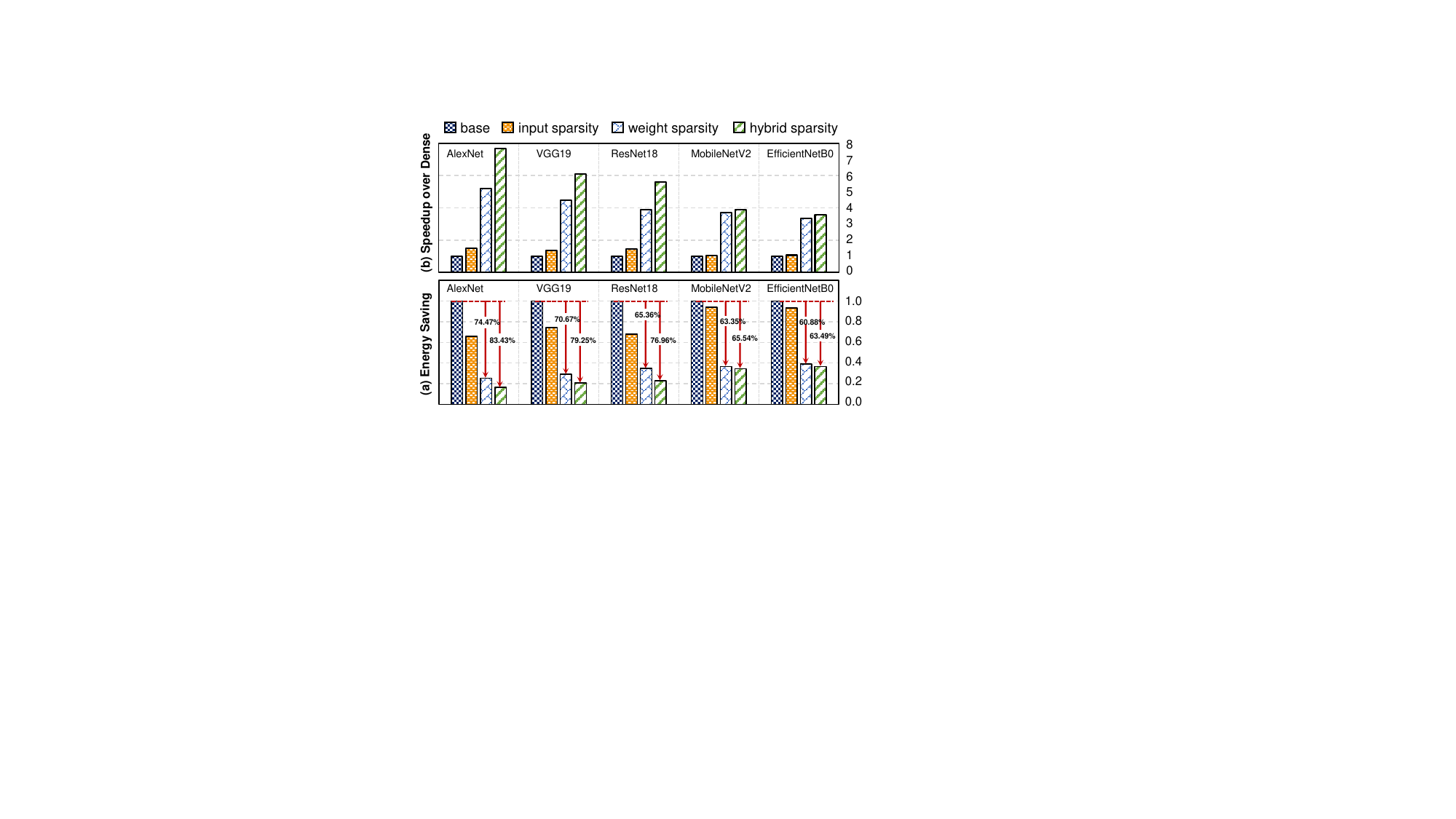}
\vspace{-18pt}
\caption{Speedup and energy savings of different sparsity exploration approaches over the dense PIM baseline.}
\label{fig8:Evaluation}
\vspace{-6pt}
\end{figure}

The accuracy comparison in Sec.~\ref{sec: Evaluation of FTA Algorithm} demonstrates that the accuracy degradation of our FTA algorithm is minimal compared to the original quantization model. 
To further explore the hardware benefits of our methods, as depicted in Fig.~\ref{fig8:Evaluation}, we present a detailed analysis of speedup and energy saving over the digital PIM baseline. 
Fig.~\ref{fig8:Evaluation}(a) illustrates the significant speedup achieved by DB-PIM across various NN models, due to its capability to exploit bit-level sparsity. 
Specifically, DB-PIM achieves a speedup of about $5.20\times$ for AlexNet and $4.46\times$ for VGG19 by utilizing weight sparsity, respectively.
Driven by high redundancy, the filter thresholds ($\phi^{th}$) of most convolutional (\texttt{Conv}) layers and fully connected (\texttt{FC}) layers in AlexNet can be set to 1 through our FTA algorithm.
Consequently, this setting allows each PIM macro within the DB-PIM to process 16 filters in parallel. 
In contrast, the acceleration in VGG19 is slightly lower because the $\phi^{th}$ for most of its \texttt{Conv} layers is set to 2, while for the \texttt{FC} layers, it remains at 1.
Despite the \texttt{FC} layer having more parameters, their computational complexity is significantly less than that of the \texttt{Conv} layers.
When considering input sparsity, the speedup for these two models increases to about $7.69\times$ and $6.10\times$, respectively.

For compact NN models, such as MobileNetV2 and EfficientNetB0, we still achieve a noteworthy acceleration of $3.90\times$ and $3.55\times$, respectively.
As detailed in \cite{tu2022sdp}, limiting sparsity to partial layers at $80\%$ in EfficientNetB0 results in just a $1.85\times$ acceleration.
This demonstrates the exceptional effectiveness of our method, particularly in EfficientNetB0 with limited redundancy.
As shown in Fig.~\ref{fig8:Evaluation}(b), the energy saving is also improved by $63.49\%$ to $83.43\%$ within five classical NN models.

\begin{table}[]
\caption{DB-PIM Area Breakdown Analysis.}
\vspace{-8pt}
\label{tab:area_breakdown}
\resizebox{0.38\textwidth}{!}{
\begin{tabular}{llr}
    \toprule
Modules                             & Area (mm$^2$)   & Breakdown      \\
    \midrule
PIM Baseline                           & 1.00809           & 87.32\%         \\
Meta-RFs                             & 0.07829           &	6.78\%         \\
Extra Post-processing Units          & 0.06259          & 5.42\%         \\
DFFs and Routing Resources               & 0.00550           & 0.48\%        \\
Input Sparsity Support              & 0.00007         & $\sim$0.00\%        \\
Total                               & 1.15453         & 100\%          \\
\bottomrule
\end{tabular}
}
\vspace{-6pt}
\end{table}

\subsection{Area Breakdown of DB-PIM}
Tab.~\ref{tab:area_breakdown} provides a detailed area breakdown analysis for DB-PIM.
The DB-PIM can be divided into the digital PIM baseline and additional logic introduced by our techniques. 
The additional logic mainly includes extra DFFs and routing resources in PIM macro, extra post-processing units with CSD-based adder tree, and meta RFs, all customized for our co-design. 
In our work, the complementary states (Q/$\overline{Q}$) in the cross-coupled structure of 6T SRAM cell represent two individual bits (the Comp. Pattern block) for parallel computations.
Hence, compared to the counterpart in \cite{yan20221}, DB-PIM requires extra storage units (DFFs) and routing resources for processing the additional information. 
The overhead incurred by these units is relatively minor, approximately $0.48\%$. 
The majority of the extra overhead stems from the additional post-processing units and meta RFs, a trade-off for enhanced parallel processing capabilities.
Typically, in standard configurations, a macro processing two 8-bit precision filters simultaneously requires only two post-processing units. 
However, our design is tailored for high-level parallel processing, allowing the concurrent computation of up to 16 filters based on the provided metadata.
This necessitates one post-processing unit for each filter, leading to an overhead that scales with the number of filters processed in parallel.
Although this approach results in a slight increase in area consumption, it significantly enhances the parallelism of the PIM array, thereby accelerating computation.

\subsection{Comparison with Prior Works}
Tab.~\ref{tab:comparison} provides a comprehensive comparison with existing state-of-the-art (SOTA) PIM-based accelerators. 
These studies represent the two predominant sparsity approaches in SRAM-PIM architectures: value-level sparsity \cite{yue202115, tu2022sdp, liu202316} and bit-level sparsity \cite{tu202316, guo2022tt}.
We mainly focus on four critical aspects: utilization, peak throughput per macro, energy efficiency, and energy efficiency per unit area.
The actual utilization $\mathcal{U}_{act}$ as illustrated in Eq.~(\ref{eq1}) in most of these studies is equal to the ratio of non-zero bits across various NN models.
Fig.~\ref{fig2:integration} shows that this ratio is extremely low for most NN models, indicating that the actual utilization in most studies is below $50\%$.
In contrast, DB-PIM effectively tackles the challenge of low utilization by employing CSD encoding, FTA algorithm, and a cross-coupled structure to store two individual bits.
Compared with previous works, $\mathcal{U}_{act}$ in DB-PIM improves up to about $3\times$. 
This innovative co-design boosts not only storage efficiency but also significantly enhances the peak throughput of each macro and energy efficiency.
The peak throughput of each macro in DB-PIM gains up to $3.14\times$ improvements compared to other works.
Meanwhile, we compare the energy efficiency and energy efficiency per unit area with previous works. 
The results show that DB-PIM reaches up to $45.20$ TOPS/W in system-level energy efficiency, surpassing most of the earlier works. 
It also shows that DB-PIM achieves the highest energy efficiency per unit area at $39.30$, much higher than previous works.

\section{Conclusion}\label{sec: Conclusion}
Bit-level sparsity holds significant potential for enhancing computational efficiency.
However, traditional digital SRAM-PIM struggles to effectively leverage randomly distributed bit-level sparsity due to their rigid crossbar structure.
Our paper presents a breakthrough in this area through an algorithm-architecture co-design framework.
DB-PIM framework adeptly utilizes unstructured bit-level sparsity in digital SRAM-PIM, overcoming the limitations of conventional designs.
The results show that DB-PIM achieves a remarkable $5.20\times$ speedup and a $74.47\%$ improvement in energy saving by utilizing unstructured weight bit sparsity.
Moreover, when combined with input bit sparsity, our framework attains even more remarkable results, with a $7.69\times$ speedup and a $83.43\%$ increase in energy efficiency. 
These results validate our co-design approach and highlight its potential to utilize the bit-level sparsity.
In the future, we aim to combine our approach with value-level sparsity to maximize the exploitation of sparsity in NN models.

\bibliographystyle{unsrt}
\bibliography{ref}

\end{document}